\documentclass[5p, preprint]{elsarticle}
\biboptions{sort&compress}
\usepackage{lineno,hyperref,isotope,mathrsfs}
\usepackage{amsmath,bbold,amssymb,epsfig,feynmp,color,ifpdf}
\usepackage{slashed,nicefrac,exscale,multirow,times,txfonts}
\usepackage{bm}
\usepackage{epstopdf}
\usepackage{booktabs}
\begin{document}
\begin{frontmatter}
\title{Microscopic Calculation of Electric Quadrupole Effective Charges in Exotic Nuclei}
\author[1,2]{Jia Liu}
\author[3]{Yong Peng}
\author[4]{Xiao-Yan Zhu}
\author[1,2]{Xiao-Hua Li\corref{cor1}}
\author[1,2]{Wen Luo}
\author[5,6]{Yi-Fei Niu}
\author[7,8]{Wen-Hui Long\corref{cor1}}
\cortext[cor1]{Corresponding authors.\\[6pt]
Xiao-Hua Li (lixiaohuaphysics@126.com)\\[6pt]
Wen-Hui Long (longwh@lzu.edu.cn)}
\address[1]{School of Nuclear Science and Technology, University of South China, Hengyang 421001, China}
\address[2]{Key Laboratory of Advanced Nuclear Energy Design and Safety, Ministry of Education, Hengyang, 421001, China}
\address[3]{School of Physics and Mechatronic Engineering, Guizhou Minzu University, Guiyang 550025, China}
\address[4]{School of Mathematics and Physics, University of South China, Hengyang 421001, China}
\address[5]{School of Physics and Astronomy, Shanghai Jiao Tong University, Key Laboratory for Particle Astrophysics and Cosmology (MoE), Shanghai 200240, China}
\address[6]{Shanghai Key Laboratory for Particle Physics and Cosmology, Shanghai 200240, China}
\address[7]{School of Nuclear Science and Technology, Lanzhou University, Lanzhou 730000, China}
\address[8]{Key Laboratory of Special Function Materials and Structure Design, Ministry of Education, Lanzhou 730000, China}

\begin{abstract}
Electric quadrupole ($E2$) effective charges are evaluated based on the self-consistent relativistic Hartree-Fock single-particle states, with core-polarization corrections resummed to all orders using the Tamm-Dancoff approximation (TDA). Configuration-interaction relativistic Hartree-Fock calculations employing the TDA effective charges well reproduce the $B(E2)$ strength for neon isotopes from stability to the neutron drip line. We find that polarization charges associated with continuum states are significantly quenched due to their extended density distributions and weak coupling to the core, underscoring the critical role of continuum effects in $E2$ transition evaluations for exotic nuclei. Moreover, the CI-RHF model predicts a suppressed $B(E2; 2^+_2 \to 0^+_2)$ in $^{30}$Ne, together with strong in-band $B(E2)$ strengths of the yrast band, suggesting the coexistence of a nearly spherical excited $0^+_2$ state and a deformed ground state within the $N=20$ "island of inversion".
\end{abstract}

\end{frontmatter}

\section{Introduction}
The advent of radioactive ion beam facilities has revolutionized the study of exotic nuclei far from stability, revealing novel structural phenomena including new magic numbers~\cite{Steppenbeck2013Nature502.207,Liu2019PRL122.072502,Enciu2022PRL129.262501,Xing2025PRL135.012501,Ong2026PRL136.092502}, the "island of inversion"~\cite{Warburton1990PRC41.1147,Motobayashi1995PLB346.9,DoornenbalPRL111.212502,Lykiardopoulou2025PRL134}, and halo structures~\cite{Tanihata2013PPNP68.215,Nakamura2014PRL113.052501,Bagchi2020PRL124.222504,Zhang2023PRC107.041303}. The reduced electric quadrupole transition probability $B(E2)$, which acts as a fingerprint of the intrinsic shape and collectivity, offers a direct and sensitive probe for tracking evolution of nuclear shape and shell structure~\cite{Heyde2011RMP83,otsuka2020RMP92}. Nevertheless, recently observed $B(E2)$ anomalies in exotic nuclei have posed a significant challenge to modern nuclear models~\cite{Imai2004PRL92.062501,Dinca2005PRC71.041302,Gustafsson2025PRL135.222501,Xu2025PRC112.L011302}.

The configuration-interaction shell model (CISM) serves as a powerful framework for investigating the low-lying spectra and electromagnetic transition properties of atomic nuclei~\cite{Caurier2005RMP77}. To reduce the dimensionality of the many-body basis, CISM calculations are typically confined to a valence space built on a frozen core. Consequently, the Hamiltonian and other observable operators must be renormalized to effectively incorporate the contributions of configurations beyond the model space. The reliability of the CISM hinges on the appropriate construction of the model space and effective operators. A prominent example is the gold-standard universal $sd$-shell Hamiltonian (USD family), derived by fitting experimental energy levels, which achieves a root-mean-square (RMS) deviation of less than 200~keV throughout the $sd$ shell~\cite{Brown2006PRC74.034315}.

Effective charges are routinely introduced to account for core-polarization effects in electric quadrupole ($E2$) transition calculations. Empirically, the proton ($\pi$) and neutron ($\nu$) effective charges, averaged over mass and valence orbitals, are determined via a least-squares fit to experimental $B(E2)$ data~\cite{Richter2008PRC78.064302}. Recently, a high-precision measurement of the $B(E2)$ strength in $^{54}$Sc, combined with existing data in the neutron-rich $pf$ shell, yielded new effective charges of $e^{\rm eff}_{\pi}=1.30(8)e$ and $e^{\rm eff}_{\nu}=0.452(7)e$~\cite{Ogunbeku2025PRL135}. These values are comparable to those used in the $sd$~\cite{Brown1988ARNPS38}, $pf$~\cite{Mizusaki1999PRC59.R1846}, and cross-shell $sdpf$ valence spaces~\cite{Caurier2014PRC90}, suggesting that "universal" effective charges can be used to describe $E2$ transitions over a broad range of atomic nuclei.

However, the assumption of constant effective charges is increasingly challenged by experiments. For instance, analysis of the $B(E2)$ strengths between high-spin states dominated by the $1f_{7/2}$ orbital in the mirror nuclei $^{51}\mathrm{Fe}$ and $^{51}\mathrm{Mn}$ yielded $pf$-shell effective charges of $e^{\rm eff}_{\pi}\sim 1.15e$ and $e^{\rm eff}_{\nu}\sim 0.80e$~\cite{Durietz2004PRL93.222501}. While adopting these effective charges improved the description of $B(E2; 0^{+}_{1} \rightarrow 2^{+}_{1})$ in neutron-rich Ti isotopes, the observed staggering remains significantly underestimated~\cite{Poves2005PRC72.047302}. Furthermore, lifetime measurements in the $N=30$ isotones $^{50}\mathrm{Ca}$ and $^{51}\mathrm{Sc}$---where the low-lying states are governed by the $\nu 2p_{3/2}$ configurations---showed that the experimental $B(E2)$ values are reproduced by the standard isoscalar core-polarization charges $e^{\rm eff}_{\pi}=1.5e$ and $e^{\rm eff}_{\nu}=0.5e$, indicating a substantial reduction of the effective neutron charge for the $\nu2p_{3/2}$ orbital relative to the $\nu 1f_{7/2}$~\cite{Valiente2009PRL102}. More recently, the isospin dependence of effective charges was isolated by comparing the half-lives of the $6^+$ and $8^+$ seniority isomers in $^{130}\mathrm{Cd}$ and $^{98}\mathrm{Cd}$, both with exceptionally pure $\pi 0g_{9/2}^{-2}$ configurations, which revealed that the isovector polarization charge decreases rapidly with increasing neutron excess, whereas the isoscalar component remains nearly constant~\cite{Jungclaus2024PRL127.222501}.

Theoretically, it is a long-standing challenge to derive effective charges for different valence spaces with a microscopic self-consistent framework. Starting from lowest-order perturbation theory, early core-polarization calculations with the Kallio–Kolltveit interaction rigorously demonstrated that effective charges depend on the initial- and final-state quantum numbers of a specific transition~\cite{Federman1969PR177}. Subsequently, the particle-hole bubble diagrams were resummed to all orders within the Tamm-Dancoff approximation (TDA) and the random-phase approximation (RPA), highlighting the importance of higher-order corrections~\cite{Siegel1970NPA145.89}. Further work by Kirson and collaborators~\cite{Kirson1971AP66,Kuo1973NPA205} showed that self-screening corrections to particle-hole and ground-state correlation vertices can drastically suppress the artificially large RPA effective charges, yielding values consistent with the TDA. In the no-core shell-model framework, Lee-Suzuki similarity transformations are used to project full-space electromagnetic operators onto truncated valence spaces. Such calculations show that phenomenological effective charges arise naturally from model-space truncation, while two-body operator components contribute marginally to $E2$ matrix elements~\cite{Navratil1997PRC55}. More recently, self-consistent Green's function approaches based on realistic nuclear interactions have yielded orbital-dependent $E2$ effective charges across the oxygen and nickel isotopic chains, and revealed a decreasing trend of neutron effective charges with increasing neutron excess~\cite{Raimondi2019PRC2.024317}.

In conventional shell-model calculations, $E2$ effective charges are typically derived from the harmonic-oscillator basis, which intrinsically describe well-bound states and therefore cannot naturally capture the long tails characteristic of weakly bound states in exotic nuclei. Hamamoto and collaborators carried out self-consistent Skyrme Hartree-Fock plus continuum RPA calculations, and obtained microscopic quadrupole polarization charges ranging from $\beta$-stable to exotic nuclei~\cite{Hamamoto1997NPA626}. Building on this work, Sagawa \textit{et al}. further constructed parametrized isospin-dependent polarization charges and incorporated them into the CISM to evaluate quadrupole moments and $B(E2)$ strengths along the carbon and neon isotopic chains~\cite{Sagawa2004PRC70.054316}. Compared with constant empirical charges, these parametrized polarization charges yield better agreement with the observed isotopic evolution of $E2$ transition strengths, though quantitative discrepancies remain.

Recently, the configuration-interaction relativistic Hartree-Fock (CI-RHF) model has been proposed to study structural properties over a wide range of nuclei~\cite{Liu2025CPC49}. Within this framework, the single-particle basis is generated from relativistic Hartree-Fock calculations, thereby capturing the spatially extended character of orbitals near the continuum threshold. Moreover, the effective interactions, as well as the core and valence-particle energies, are derived from a unified phenomenological Lagrangian for different model spaces. The CI-RHF model has been shown to reproduce well the ground-state and low-lying spectral properties of nuclei in the $sd$ and $pf$ shells~\cite{Peng2025CPC49.064112,liu2026CPC50.091002}. However, the $B(E2)$ strengths in neutron-rich Ne isotopes are significantly overestimated when using constant effective charges, indicating the importance of orbital and isospin dependence of effective charges \cite{Liu2025CPC49}. In this work, effective $E2$ charges are evaluated self-consistently in the relativistic Hartree-Fock basis, with core-polarization corrections summed to all orders via the Tamm-Dancoff approximation. These effective charges are then employed in CI-RHF calculations to obtain the $B(E2)$ strengths in exotic nuclei, allowing us to further explore the interplay between electric quadrupole transitions and continuum effects.

\section{Theoretical framework}
Within Rayleigh-Schr\"odinger perturbation theory, an energy-independent effective Hamiltonian is constructed via a similarity transformation~\cite{Kuo1973NPA205}:
\begin{equation}
\mathscr{H} = e^{-\chi} H e^{\chi}, \qquad Q\chi P = \chi,
\end{equation}
where $\chi$ is the correlation operator that maps the model space $P$ to the complementary space $Q = 1 - P$. Imposing the decoupling condition $Q\mathscr{H}P = 0$ yields
\begin{equation}\label{eq:decoupling_eq}
QHP - \chi HP + QH\chi P - \chi H\chi P = 0.
\end{equation}
The energy expectation value is then determined entirely within the model space:
\begin{equation}
E = \langle \Psi | e^{\chi} \mathscr{H} e^{-\chi} | \Psi \rangle = \langle \Psi | P H^{\mathrm{eff}} P | \Psi \rangle,
\end{equation}
with the effective Hamiltonian
\begin{equation}
H^{\mathrm{eff}} = PH_{0}P + PVP + PH\chi P,
\end{equation}
where $H_{0}$ is the unperturbed Hamiltonian and $V$ is the residual interaction. The nonlinear decoupling equation~\eqref{eq:decoupling_eq} can be solved iteratively via the extended Krenciglowa--Kuo (EKK) scheme~\cite{Krenciglowa1975NPA240}, in which $\chi$ is expanded about the starting energy $E_{0}$:
\begin{equation}\label{eq:chi_expansion}
\chi_{n} = \sum_{k=0}^{\infty} (-1)^{k} \frac{Q}{(E_{0}-QHQ)^{k+1}} \, V P \, (H^{\mathrm{eff}}_{n-1}-E_{0})^{k},
\end{equation}
where $n$ is the iteration index. The effective Hamiltonian at the $n$-th iteration reads:
\begin{equation}\label{eq:Heff_n}
H^{\mathrm{eff}}_{n} = PH_{0}P + \sum_{k=0}^{\infty} \frac{1}{k!}\frac{d^{k}\hat{Q}(E_{0})}{dE_{0}^{k}} (H^{\mathrm{eff}}_{n-1}-E_{0})^{k},
\end{equation}
where the $\hat{Q}$-box is defined as
\begin{equation}
\hat{Q}(E) = PVP + PV\frac{Q}{E - H}VP.
\end{equation}
Since the resolvent $Q/(E-H)$ in the $\hat{Q}$-box cannot be evaluated directly, it is further expanded about $H_{0}$:
\begin{equation}\label{eq:Q_expansion}
    \frac{Q}{E-H} 
    = \frac{Q}{E-H_{0}} + \frac{Q}{E-H_{0}}\,V\,\frac{Q}{E-H_{0}} + \cdots.
\end{equation}
It is noted that both $\chi_{n}$ and $H^{\mathrm{eff}}_{n}$ are expressed as Taylor expansions at the starting energy $E_{0}$, so that a shift of $E_{0}$ merely redistributes contributions among different orders and the converged result becomes independent of the starting energy once all folding terms ($k\geq 1$) are summed.

Once $\chi$ is obtained, the full wave function is expressed as
\begin{equation}
|\Psi\rangle = (1 + \chi)\, P|\Psi\rangle.
\end{equation}
The effective operator $O^{\rm eff}$ is introduced to account for the physics excluded by the model-space truncation:
\begin{equation}\label{eq:eff_def}
\langle \Psi | O | \Psi \rangle = \langle \Psi | P O^{\rm eff} P | \Psi \rangle,
\end{equation}
and takes the form:
\begin{equation}\label{eq:eff_op_general}
O^{\rm eff} = POP + PO\chi P + P\chi^{\dagger} O P + P\chi^{\dagger} O \chi P + \cdots,
\end{equation}
where the first term is the bare operator and the second and third terms correspond to the first-order correction in $\chi$, with the higher-order corrections in $\chi$ neglected in the present work.

The electric quadrupole transition probability $B(E2)$ between initial state $|\Psi_{i}\rangle$ and final state $|\Psi_{f}\rangle$ is defined as
\begin{equation}\label{eq:BE2_def}
    B(E2; \Psi_{i}\rightarrow \Psi_{f}) 
    = \frac{1}{2J_{i}+1}\,\big|\langle \Psi_{f} || O(E2) || \Psi_{i} \rangle\big|^{2},
\end{equation}
where the double-bar notation denotes the reduced matrix element. The electric quadrupole transition operator, $O(E2)$, is a one-body tensor of rank $L=2$, and its bare matrix element between single-particle states $a$ and $b$ reads
\begin{equation}
    O_{ab}(E2) \equiv e\, M(E2; b\rightarrow a) 
    = e\,\langle a | r^{2} \bm{Y}_{2}| b \rangle,
\end{equation}
where $e$ is the bare nucleon charge and $M(E2; b\rightarrow a)$ denotes the quadrupole matrix element from $b$ to $a$. Although the effective operator in Eq.~\eqref{eq:eff_op_general} contains many-body terms of rank $L=2$, these contribute marginally to $E2$ transitions~\cite{Navratil1997PRC55}. In the present work, only the one-body terms are retained, and the renormalized contributions can be absorbed into the effective charge:
\begin{equation}\label{eq:eff_charge_def}
O^{\rm eff}_{ab}(E2) \equiv e^{\rm eff}_{ab}\, M(E2; b\rightarrow a),
\end{equation}
where $e^{\rm eff}_{ab}$ depends on the valence single-particle states $a$ and $b$.

\begin{figure}[htbp]\setlength{\abovecaptionskip}{0.0em}
  \centering
  \includegraphics[width=1.00\linewidth]{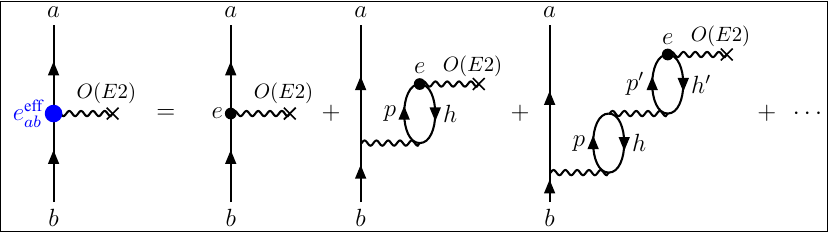}
  \caption{Diagrammatic representation of the perturbative expansion of the effective $E2$ charge $e_{\rm eff}^{ab}$ in the Tamm--Dancoff approximation. The left-hand side denotes the full effective operator. The right-hand side decomposes into: (i)~the zero-order term with bare charge $e$, (ii)~the first-order core polarization with a single particle-hole bubble, (iii)~the second-order correction with two successive particle-hole bubbles, and higher-order terms indicated by the ellipsis. The $O(E2)$ denotes the electric quadrupole operator insertion.}\label{fig:tda}
\end{figure}

Using the resolvent expansion of Eq.~\eqref{eq:Q_expansion} to evaluate $\chi$ and retaining all-order forward-going particle-hole bubble diagrams within the Tamm--Dancoff approximation (TDA), the matrix element of the effective operator $O^{\rm eff}_{ab}(E2)$ and the effective $E2$ charges $e^{\rm eff}_{ab}$ are derived. As illustrated in Fig.~\ref{fig:tda}, the single-bubble diagram represents the core-polarization correction arising from the coupling of the valence nucleon to particle-hole excitations:
\begin{equation}\label{eq:cp_first}
O_{ab}^{(1)}(E2) = \sum_{ph} \frac{\bar{V}_{apbh}\,O_{hp}(E2)}{R(ph)},
\end{equation}
where $\bar{V}_{apbh}$ denotes the antisymmetrized two-body interaction matrix element, and $p$ and $h$ represent particle and hole states, respectively. The energy denominator reads $R(ph) \equiv E_{0} - (\varepsilon_{a} + \varepsilon_{p} - \varepsilon_{h})$. Since the valence particle energies are nearly degenerate for $E2$ transitions within a single harmonic-oscillator shell, the starting energy is approximated as $E_{0} \approx \varepsilon_{a} \approx \varepsilon_{b}$, which simplifies the denominator to $R(ph) \approx \varepsilon_{h} - \varepsilon_{p}$. Furthermore, the contributions from the folding terms ($k \geq 1$) in Eq.~\eqref{eq:chi_expansion}, which correct the valence particle energies, are omitted in the present calculation.

The second-order and higher-order terms in Fig.~\ref{fig:tda} involve successive particle-hole excitations, where each additional bubble introduces a particle-hole matrix element $\bar{V}$ and corresponding energy denominator $R$. Summing the forward-going bubble chains to all orders yields
\begin{equation}\label{eq:cp_higher}
    O_{ab}^{\rm TDA}(E2)
    = \sum_{php'h'} \frac{\bar{V}_{apbh}}{R(ph)}
    \left[\delta_{pp'}\delta_{hh'} + \frac{\bar{V}^{\rm eff}_{hp'ph'}}{R(p'h')}\right]
    O_{h'p'}(E2),
\end{equation}
where $\bar{V}^{\rm eff}$ is the renormalized interaction that incorporates the infinite particle-hole bubble diagrams via the self-consistent recursion:
\begin{equation}\label{eq:Veff}
    \bar{V}^{\rm eff}_{hp'ph'} = \bar{V}_{hp'ph'}
    + \sum_{p''h''} \frac{\bar{V}_{hp''ph''}\,\bar{V}^{\rm eff}_{h''p'p''h'}}{R(p''h'')}.
\end{equation}
Collecting the bare term and the all-order bubble resummation, the TDA effective charge reads:
\begin{equation}\label{eq:resum}
e^{\rm TDA}_{ab} = \frac{O_{ab}(E2) + O_{ab}^{\rm TDA}(E2)}{M(E2; b\rightarrow a)}.
\end{equation}
For comparison, the first-order core-polarization (CP) effective charge retains only the leading correction:
\begin{equation}\label{eq:ecp}
e^{\rm CP}_{ab} = \frac{O_{ab}(E2) + O_{ab}^{(1)}(E2)}{M(E2; b\rightarrow a)}.
\end{equation}

It is worth noting that the complete first-order correction in $\chi$ also incorporates the contribution where the $E2$ operator acts prior to the residual interaction. This corresponds to the term $P\chi^{\dagger}OP$ in Eq.~\eqref{eq:eff_op_general} and is evaluated by interchanging the indices $a \leftrightarrow b$ in the particle-hole matrix elements. In the numerical implementation, both one-body transition and two-body interaction matrix elements are evaluated in the angular momentum-coupled representation with the single-particle basis obtained from self-consistent relativistic Hartree-Fock calculations.

\section{Results and discussion} \label{sec:results}
Neon isotopes span from $N=8$ to $N=20$, allowing for a systematic study of the evolution of electric quadrupole collectivity and the interplay between continuum coupling and core-polarization effects. In this work, we present CI-RHF calculations of excitation spectra and $B(E2)$ transition strengths for the even-even neon isotopes $^{18\text{--}32}$Ne. The model space consists of the $sd$-shell valence orbitals built upon a $^{16}$O closed-shell core for $^{18-28}$Ne. To account for cross-shell excitations in the $N=20$ "island of inversion", the valence space is extended to include the neutron $\nu1f_{7/2}$ and $\nu2p_{3/2}$ orbitals for $^{30,32}$Ne.

\begin{figure}[htbp]\setlength{\abovecaptionskip}{0.0em}
  \centering
  \includegraphics[width=1.0\linewidth]{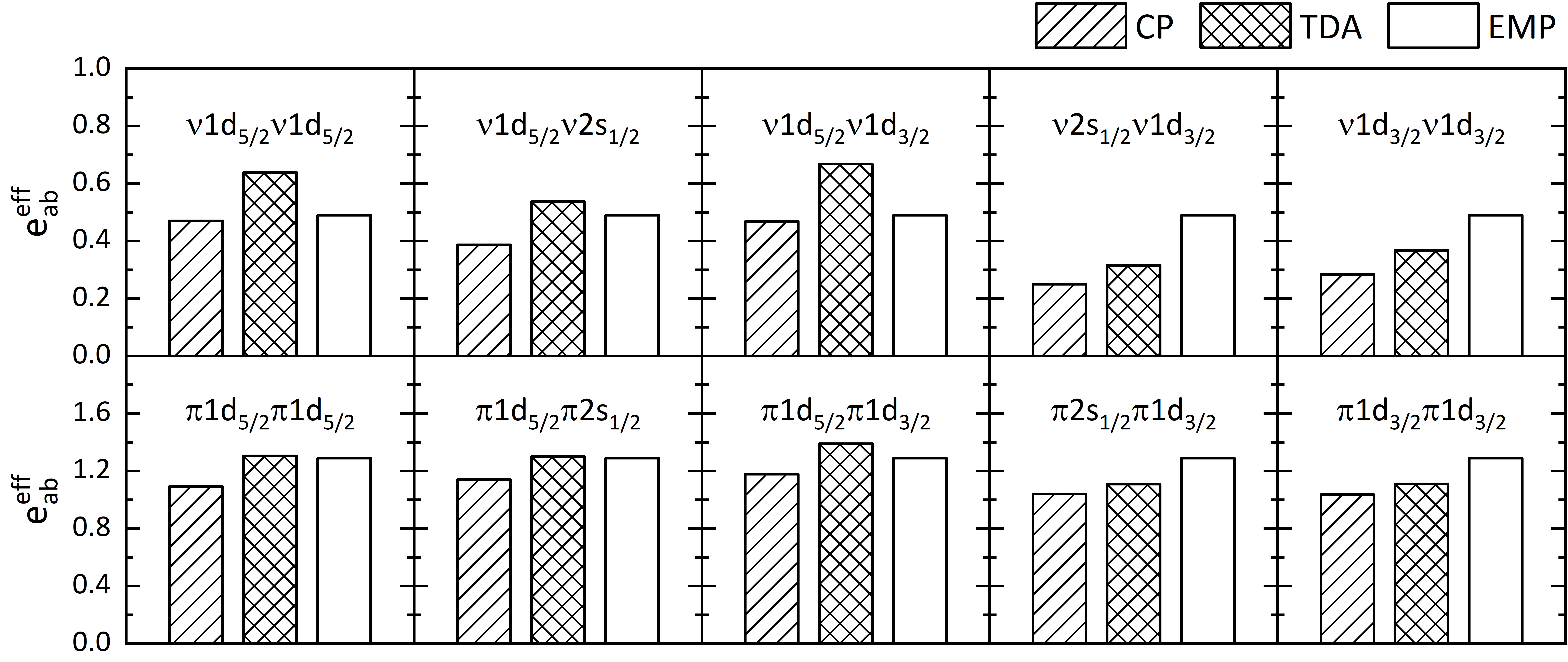}
  \caption{$sd$-shell effective charges $e^{\rm eff}_{ab}$ of $^{20}$Ne. Theoretical results are obtained with the PKA1 Lagrangian: the first-order core-polarization effective charge $e^{\rm CP}_{ab}$ (CP, diagonal hatching) and the TDA effective charge $e^{\rm TDA}_{ab}$ (TDA, cross-hatching). The empirical values $e^{\rm eff}_{\pi}=1.29e$ and $e^{\rm eff}_{\nu}=0.49e$ are shown as open bars for comparison.}\label{fig:comp}
\end{figure}

Figure~\ref{fig:comp} displays the calculated effective charges of $^{20}$Ne, compared with the empirical values averaged over $sd$-shell orbits and mass number~\cite{Brown1988ARNPS38}. The theoretical results are obtained using the PKA1 Lagrangian~\cite{Long2007PRC76.034314}, with core-polarization corrections calculated to first order and to all orders in the TDA. For all one-body $E2$ transition channels, the TDA effective charges are systematically larger than their first-order counterparts. As shown in Eq.~\eqref{eq:cp_higher}, all higher-order bubble corrections are resummed into a dressed factor $\bar{V}^{\text{eff}}/R$ through the recursion of Eq.~(\ref{eq:Veff}). Since the renormalized interaction $V^{\rm eff}$ is attractive for the isoscalar quadrupole resonance, this factor is positive and thus amplifies the core-polarization correction.

Moreover, the polarization charges exhibit a pronounced isospin asymmetry: the averaged neutron polarization charge $e^{\rm pol}_{\nu}=0.51e$ is approximately twice the proton value $e^{\rm pol}_{\pi}=0.24e$. A decomposition into isoscalar and isovector components, $e^{\rm pol}_{\rm IS}=(e^{\rm pol}_{\nu}+e^{\rm pol}_{\pi})/2=0.38e$ and $e^{\rm pol}_{\rm IV}=(e^{\rm pol}_{\nu}-e^{\rm pol}_{\pi})/2=0.14e$, reveals that the quadrupole response of the $^{16}$O core is predominantly isoscalar, with neutrons and protons oscillating in phase. The residual isovector component, corresponding to out-of-phase oscillations of neutrons and protons, further enhances the neutron polarization charge relative to the proton one~\cite{Hamamoto1997NPA626,Sagawa2004PRC70.054316}.

The calculated $sd$-shell effective charges show clear orbital dependence, where the effective charge associated with the initial $\nu1d_{5/2}$ orbital exceeds those of the $\nu2s_{1/2}$ and $\nu1d_{3/2}$ orbitals by approximately $0.3e$, a similar trend being observed for protons albeit with a smaller magnitude. It is noted that the orbit-averaged TDA effective charges for $^{20}$Ne are $e^{\rm eff}_{\pi}=1.24e$ and $e^{\rm eff}_{\nu}=0.51e$, in reasonable agreement with the empirical values for the $sd$ shell~\cite{Brown1988ARNPS38}. However, as illustrated below for $^{30}$Ne, the constant effective charges fails to describe $B(E2)$ strengths for transitions dominated by different valence orbits within the same nucleus.

\begin{figure}[htbp]\setlength{\abovecaptionskip}{0.0em}
  \centering
  \includegraphics[width=0.90\linewidth]{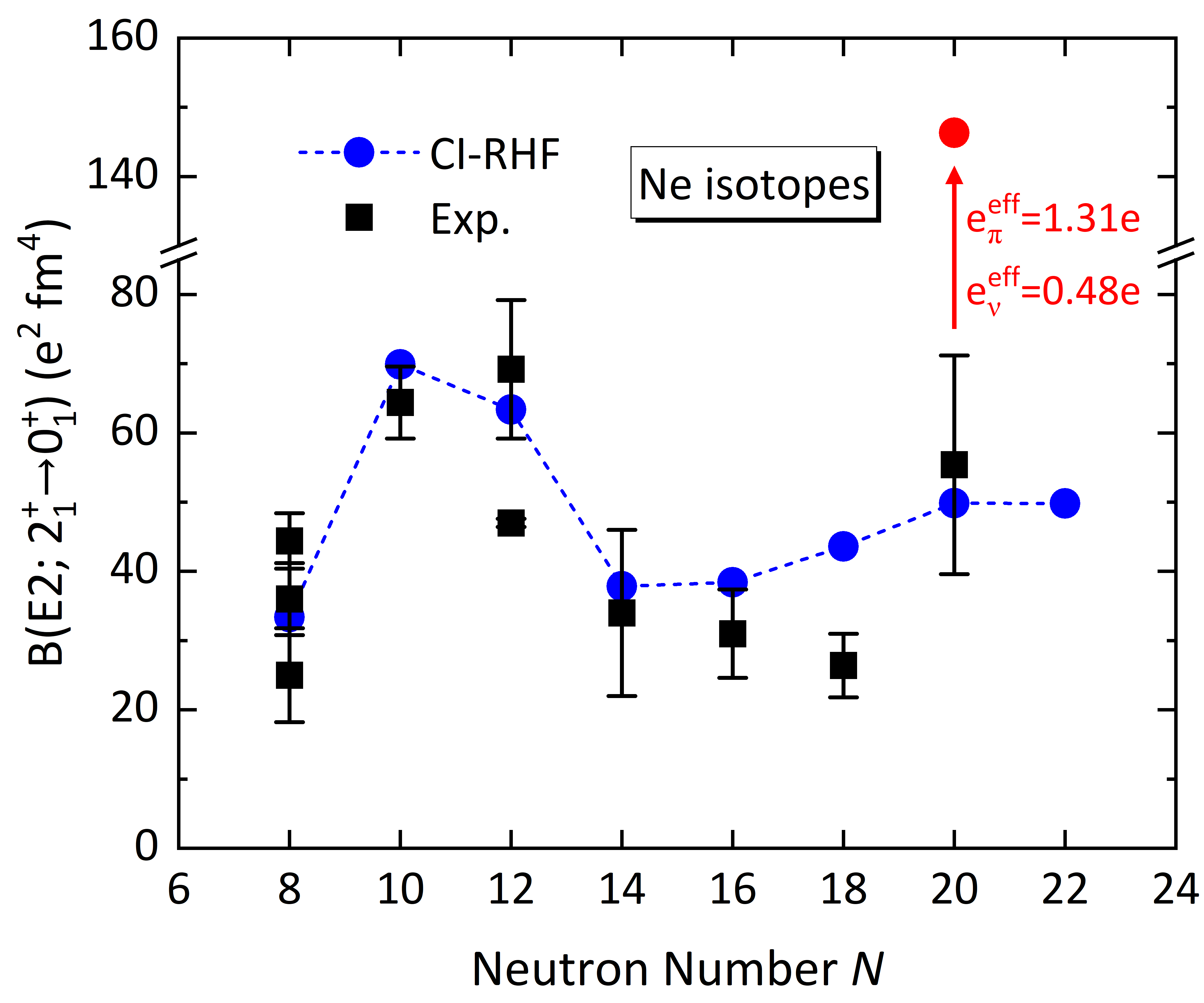}
  \caption{(Color online) $B(E2; 2_1^+\to 0_1^+)$ values of the even-even Ne isotopes as a function of neutron number $N$. The black squares and blue circles denote the experimental data and the CI-RHF calculations with the orbital-dependent TDA effective charges, respectively. The red circle for $^{30}$Ne represents the value obtained with the constant effective charges $e^{\rm eff}_{\pi}=1.31e$ and $e^{\rm eff}_{\nu}=0.48e$ averaged over the $sd$-shell orbitals.}\label{fig:be2_ne}
\end{figure}

Using the derived TDA effective charges, we calculate the $B(E2; 2_1^+\to 0_1^+)$ values of the even-even Ne isotopes within the CI-RHF model and compare them with the available experimental data in Fig.~\ref{fig:be2_ne}. The overall trend of the $B(E2)$ strength from $N=8$ to $N=22$ is well reproduced, despite an overestimate for $^{28}$Ne, as also found in Monte Carlo shell model calculations~\cite{Utsuno1999PRC60}. An earlier experimental measurement yielded an even larger $B(E2)$ for $^{28}$Ne~\cite{Pritychenko1999PLB461}, and the discrepancy between the two data sets remains unresolved. The evolution of $B(E2)$ strengths along the Ne isotopic chain provides a sensitive probe of nuclear structural changes. The calculated small $B(E2)$ value for $^{18}$Ne is attributed to a near-spherical ground-state configuration, suggesting an underlying $N=8$ shell closure. In contrast, the pronounced enhancements observed at $N=10$ and $12$ identify $^{20,22}$Ne as well-deformed nuclei with strong quadrupole collectivity~\cite{Marinova2011PRC84.034313}. The subsequent decrease in $B(E2)$ for $^{24}$Ne and $^{26}$Ne signals a return toward sphericity, driven by the $N=14$ and $N=16$ subshell structures~\cite{Ozawa2000PRL84.5493,Kanungo2009PRL102.152501}. Finally, the rising $B(E2)$ strength approaching $N=20$ indicates the collapse of the $N=20$ shell gap. If the valence space is restricted to the $sd$ shell, i.e., the $N=20$ closed-shell configuration, the calculated $B(E2)$ for $^{30}$Ne is reduced to 27.9~e$^2$fm$^4$, comparable to that of $^{18}$Ne. The enhancement of the $B(E2)$ for $^{30}$Ne thus arises from neutron cross-shell excitations from the $sd$ into the $pf$ shell, consistent with the picture of the $N=20$ "island of inversion" established by experiment and shell-model studies~\cite{Motobayashi1995PLB346.9,Caurier2014PRC90}.

\begin{figure}[htbp]\setlength{\abovecaptionskip}{0.0em}
  \centering
  \includegraphics[width=0.90\linewidth]{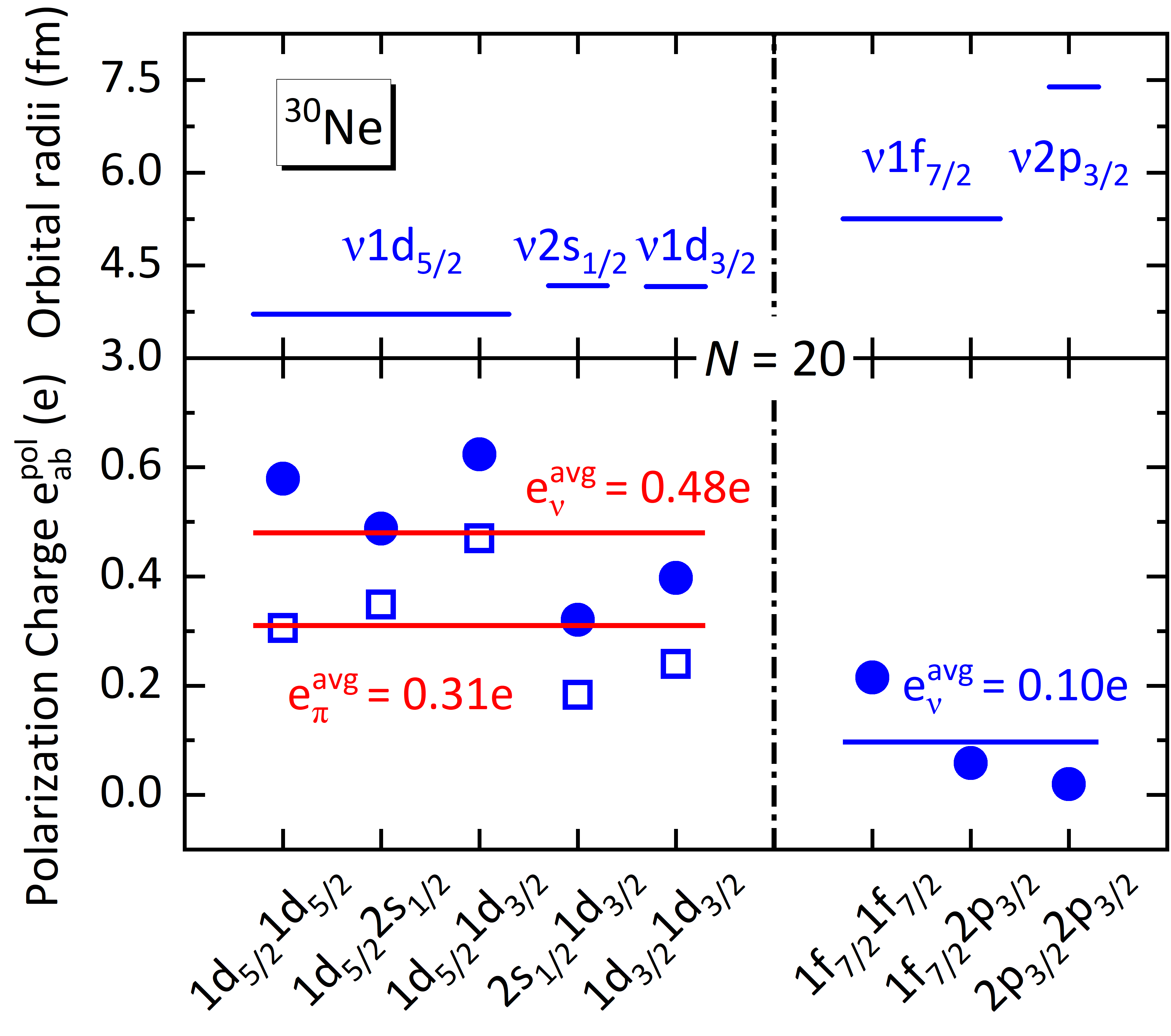}
  \caption{(Color online) Neutron (blue filled circles) and proton (open blue squares) TDA polarization charges $e^{\rm pol}_{ab}$ in $^{30}$Ne (lower panel), and the RMS radii of the corresponding neutron orbitals $a$ (upper panel). The horizontal lines in the lower panel denote orbital-averaged polarization charges within the $sd$ shell ($e^{\rm avg}_{\pi}=0.31e; e^{\rm avg}_{\nu}=0.48e$) and $pf$ shell ($e^{\rm avg}_{\nu}=0.10e$).}\label{fig:ec_ne}
\end{figure}

The present theoretical calculations quantitatively reproduce the experimental $B(E2)$ strengths, validating the self-consistent, orbital- and isospin-dependent TDA effective charges of Eq.~\eqref{eq:resum} in both single- and multi-shell valence spaces. To further elucidate the evolution of effective charges from stable to exotic nuclei, the lower panel of Fig.~\ref{fig:ec_ne} presents the polarization charges for $^{30}$Ne. Compared with $^{20}$Ne, the orbit-averaged $sd$-shell isoscalar polarization charge remains nearly unchanged at $e^{\rm avg}_{\rm IS} = 0.39e$, while the isovector component drops to $e^{\rm avg}_{\rm IV} = 0.09e$, comparable to that in $^{132}$Sn \cite{Jungclaus2024PRL127.222501}. Furthermore, the polarization charges for the $\nu1f_{7/2}$ and $\nu2p_{3/2}$ orbitals beyond the $sd$ shell are strongly quenched to an average of $e^{\rm avg}_{\nu}=0.10e$, indicating a significantly enhanced orbital dependence towards the neutron drip-line.

According to Eq.~\eqref{eq:cp_first}, the orbital dependence originates from the coupling between valence nucleons and collective quadrupole vibrations of the core~\cite{Raimondi2019PRC2.024317}. As illustrated in the upper panel of Fig.~\ref{fig:ec_ne}, the $\nu1f_{7/2}$ and $\nu2p_{3/2}$ orbitals exhibit significantly larger rms radii than those within the $sd$ shell; such extended spatial distributions result in a weak overlap with the inner core density, thereby reducing core-polarization corrections. Consequently, the polarization charges decrease progressively with increasing orbital radius, approaching the bare nucleon charges for continuum states.

In the harmonic-oscillator basis, the so-called universal empirical charges $e^{\rm eff}_{\pi} \approx 1.3e$ and $e^{\rm eff}_{\nu} \approx 0.5e$ are routinely adopted for $B(E2)$ calculations in both single- and multi-shell valence spaces~\cite{Ogunbeku2025PRL135}. However, within a self-consistent Hartree-Fock single-particle basis, such pronounced orbital dependence cannot be captured by the smooth $(N,Z)$ systematics of empirical effective charges~\cite{Sagawa2004PRC70.054316}. When constant effective charges (averaged over the $sd$ shell) are used to evaluate the $B(E2)$ strength in $^{30}$Ne within the $sd$-$pf$ valence space, the transition strength is substantially overestimated (blue circle in Fig.~\ref{fig:be2_ne}), underscoring the critical role of continuum effects in $B(E2)$ evaluations for exotic nuclei. Recently, one-neutron removal cross-section measurements have identified a $p$-wave halo component in $^{31}$Ne~\cite{Nakamura2014PRL113.052501}, motivating further exploration of the decoupled character between halo orbitals and the core via $B(E2)$ strengths in halo nuclei~\cite{SUN2021SB}.

\begin{figure}[htbp]\setlength{\abovecaptionskip}{0.0em}
  \centering
  \includegraphics[width=0.90\linewidth]{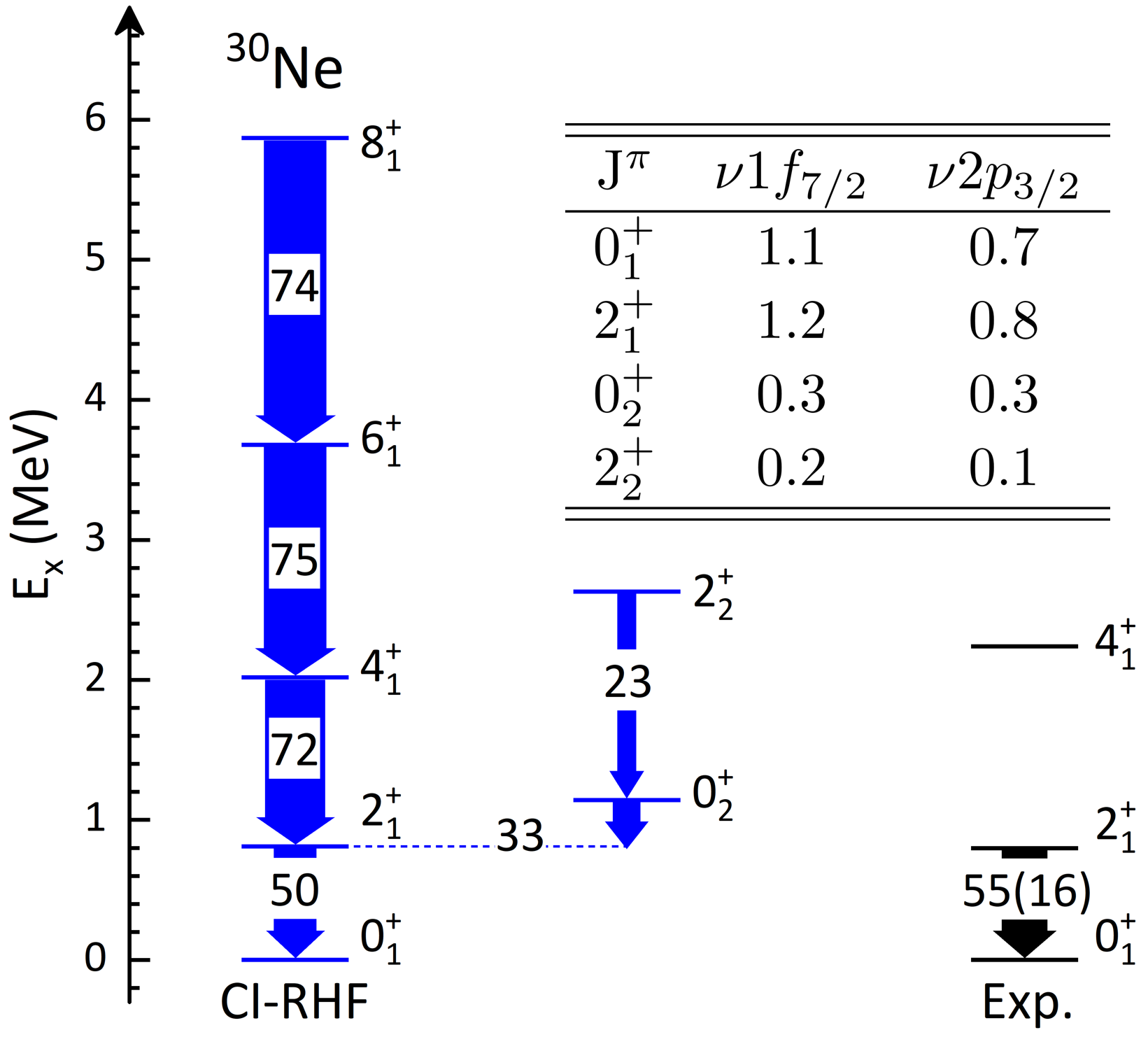}
  \caption{(Color online) Low-lying spectrum of $^{30}$Ne calculated with the CI-RHF model (left) in comparison with the available experimental levels (right)~\cite{Yanagisawa2003PLB566}. The arrows indicate the calculated $B(E2)$ values in e$^2$fm$^4$. The inset table lists the neutron occupation numbers of the $\nu1f_{7/2}$ and $\nu2p_{3/2}$ intruder orbitals for the low-lying states.}\label{fig:ne30}
\end{figure}

Shape coexistence has been intensively studied in the vicinity of the "island of inversion". The coexistence of a deformed, intruder-dominated ground-state band with a nearly spherical excited state $0^{+}_{2}$ was suggested in $^{32}$Mg~\cite{Wimmer2010PRL105,Motobayashi1995PLB346.9}. However, subsequent shell-model calculations using the SDPF-U-MIX Hamiltonian indicate that the excited $0^+_2$ state contains significant spherical and superdeformed components, rendering the shape coexistence picture in the $^{32}$Mg more complex \cite{Caurier2014PRC90.014302}. Similarly, CI-RHF calculations predict a low-lying excited $0^+_2$ state in $^{30}$Ne, as shown in Fig.~\ref{fig:ne30}. The ground-state band exhibits strong in-band transition with $B(E2)$ strengths of 50 -- 75~e$^2$fm$^4$, consistent with the large deformation deduced from intermediate-energy inelastic scattering~\cite{Doornenbal2016PRC93}. This picture is substantiated by the calculated configurations in the inset table, where the yrast band involves significant cross-shell excitations, with approximately two neutrons promoted to the $\nu1f_{7/2}$ and $\nu2p_{3/2}$ orbitals. In contrast, the $0_2^+$ and $2_2^+$ states are dominated by normal $sd$-shell configurations with only minor intruder admixtures, and this near-spherical structure results in a much weaker $B(E2; 2_2^+\to 0_2^+)=23$~e$^2$fm$^4$, comparable to that of $^{18}$Ne. The distinctly different $E2$ transition strengths of the two bands thus provide a clear spectroscopic signature of the coexistence between a deformed ground state and a near-spherical excited $0^+_2$ state. The inter-band transition strength $B(E2; 0_2^+\to 2_1^+)=33$~e$^2$fm$^4$ further indicates relatively weak configuration-mixing between the yrast band and the excited $0^+_2$ state.

\section*{Summary}
In summary, orbital-dependent $E2$ effective charges have been derived within the configuration-interaction relativistic Hartree-Fock (CI-RHF) framework, with core-polarization corrections resummed to all orders in the Tamm-Dancoff approximation. The orbital-averaged TDA effective charges of $^{20}$Ne are $e^{\rm eff}_{\pi}=1.24e$ and $e^{\rm eff}_{\nu}=0.51e$, in reasonable agreement with empirical values in $sd$ shell, while a pronounced orbital dependence emerges toward the drip line, where the extended spatial density distributions of loosely bound orbitals weaken the coupling to the core and lead to nearly quenched polarization charges. The present findings suggest that continuum effects are crucial for reliable $E2$ transition predictions in exotic nuclei. With the self-consistent TDA effective charges, the CI-RHF calculations reproduce the $B(E2)$ strengths for neon isotopes from stability toward the neutron drip line. Furthermore, the CI-RHF predict a strongly suppressed $B(E2; 2^+_2 \to 0^+_2)$ in $^{30}$Ne, together with the enhanced in-band $B(E2)$ strengths of the yrast band, providing a key spectroscopic signature for the coexistence of a nearly spherical excited $0^+_2$ state and a deformed ground state.

\section*{Acknowledgements}
This work was supported by the National Natural Science Foundation of China under Grant Nos.~12547177, 12375113 and 12035011.

\bibliographystyle{elsarticle-num}
\bibliography{reference}

\end{document}